\documentclass[preprint,showpacs,preprintnumbers,amsmath,amssymb,floats]{revtex4-1}
\usepackage{blindtext}
\usepackage{graphicx}
\usepackage[english]{babel}
\usepackage{amsmath}
\usepackage{amssymb}
\usepackage{color}

\newcommand{\be}{\begin{eqnarray}}
\newcommand{\ee}{\end{eqnarray}}
\usepackage[toc,page]{appendix}

\begin{document}
\title{Dark Fermions in Fluctuating Valence Insulators}
\author{C. M. Varma}
\affiliation{University of California, Berkeley, CA. 94720\\
\thanks{Visiting Scholar} 
University of California, Riverside, CA. 92521} 
\thanks{Emeritus}
\date{\today}
\begin{abstract} 
A fluctuating-valence impurity in a metal is quantum-critical unlike a Kondo impurity which has the properties of a local Fermi-liquid. A systematic theory  for the fluctuating-valence lattice is  constructed, based on the  hybridization and pairing of itinerant d-orbitals with localized f-orbitals both of  which are essential parts of the solution of the impurity problem.  It also uses the fact that the single-particle excitations at the Fermi-surface in any dimension can  be written as orthogonal Majoranas and those with linear departures from the Fermi-surface as linear combination of bare particles and holes with the same spin. The calculations on the lattice give four spin-degenerate one-particle excitations of fractionalized fermions; two sets disperse across the chemical potential and the other two have gaps. The former are shown to be dark to any linear electro-magnetic probes of their charge and spin and observable only through probes of their free-energy such as a Fermi-liquid specific heat and magneto-oscillations characteristic of a Fermi-surface but without a Zeeman splitting. The excitations with the gaps behave as in insulators but with renormalized amplitudes. The superfluid density is zero. A magnetic field $H$ turns the insulator to a metal with a singularity in magnetization proportional to $\sqrt{H - H_c}$, with $H_c$ related to the gap. Beyond $H_c$,  the usual Zeeman splitting appears in the magneto-oscillations. The properties and predictions are compared to the  momentous  recent discoveries in fluctuating-valence insulators. Similar excitations may be expected in transition metal chalcogenide layers at fluctuating-valence, and quite likely for  Kagome lattices, and twisted multi-layer graphene near specific fillings. 
\end{abstract}

\maketitle

\section{Introduction}

 A truly remarkable discovery of the past decade are the magneto-oscillations periodic in $1/H$ 
characteristic of a Fermi-surface
in  fluctuating mixed-valence insulators SmB$_6$ and YbB$_{12}$ \cite{Sebastian2015SmB6, Li2018YbB12, Sebastian2018YbB12,  SebastianSmB62020}. Recent measurements \cite{Singleton2022} show that the change of the amplitude 
of the oscillations with temperature follows the Lifshitz-Kosevich formula which relies only on having  excitations with a Fermi-distribution.  
The fermion mass found in such measurements and the size of the Fermi-surface from the period of oscillations,  using the 
Onsager flux quantization condition, gives the coefficient of the linear in temperature specific heat (and associated
thermal conductivity) in good agreement with  measurements \cite{Sato2019} as for metals. On the other hand dc resistivity, optical conductivity  \cite{Optics_Armitage_2017}  and neutron scattering \cite{YbB12Neutrons2007} \cite{Optics_Armitage_2017} show that the density of particle-hole excitations is $0$ at low frequencies and temperatures.  %ARPES  and tunneling measurements show a more complicated picture of single-particle excitations which we will discuss further below \cite{YbB12_2015_HybGap}.  
There are other rather unique features of the recent 
results \cite{Singleton2021, Singleton_aps_2023, RevAllen2020}, which will be addressed below. Older experiments are reviewed in \cite{V_RMP_MV, Riseborough2000}.

An  essentially exact result, obtained by Wilson's numerical renormalization group and a bosonization solution,  is that unlike the Kondo impurity in a metal, the correct minimal model for a fluctuating-mixed-valence impurity has logarithmic low 
energy singularities in its local charge and spin-susceptibility as well as in its local pairing susceptibility \cite{PerakisVRuck, Sire_V_R_G}. 
The 
excitations in  the same model  can be expressed as Majoranas \cite{Varma2020_Majorana} which do not couple linearly to electromagnetic fields. This was extended to
construct a theory for the lattice \cite{Varma2020_Majorana}. But the nature of the non-local  particles on which the periodic Bloch conditions are imposed in such a theory is physically obscure, which makes that  procedure of extension to the lattice unclear. A direct approach to the problem in the lattice is used here, based on an ansatz also motivated by the solution of the single impurity problem. The consistency of the ansatz is proven and leads to a state with close correspondence to the experiments and several predictions. 

This paper is organized thus: The model for the mixed-valence lattice is presented in Sec. II, where first the difference of the Kondo impurity in a metal from a mixed-valence impurity is recalled. Fig. (\ref{Fig:KondoMV}) clarifies this further. In the next Sec. III, a pairing ansatz for the lattice problem for local $f$ and $d$ orbitals is described through a reduced Hamiltonian at centers of mass momenta $2{\bf k}_F$, where ${\bf k}_F$ are the Fermi-surface. This allows the use of a Hamiltonian for lattice fermions expressible for small departure ${\bf q}$ from ${\bf k}_F$ on the basis of fermions which are equal linear combination of particles and holes which at ${\bf k}_F$ are purely real. The single-particle excitations determined variationally are used in combination with the condition for stable mixed-valence to derive the parameters of the excitation spectra in Sec. IV. In Sec V, these results are used to show that magneto-oscillations and specific heat with all the properties of the usual metal but without the Zeeman splitting follow naturally. On the other hand, these excitations are dark to electromagnetic probes including dc conductivity, and invisible also in ultra-sound and NMR experiments. They should be found in tunneling experiments but with much reduced amplitude. The superfluid density is also shown to be $0$.  In the concluding section,  some of the theoretical issues and other possible applications are mentioned.

\section{The Model}

 The physical reasons for the difference of a fluctuating-valence impurity from a Kondo impurity lie in the requirement that both valences satisfy the Friedel screening condition or local charge neutrality and have been already described \cite{Varma_Heine1975, Haldane1977_mixedval, CMVcorrins}. This requires at least one additional interaction parameter in the model in order to enforce charge neutrality for both charge states of the impurity. Mixed-valence requires equality of the chemical potential for the two local charge states. So, at mixed-valence, there are then two competing fluctuating channels, the valence fluctuations of the f-states with a corresponding local charge in the uncorrelated d-orbital and the spin-fluctuations as in the Kondo problem for the magnetic f-states.  Both channels have logarithmic singularities about the unstable ultra-violet fixed point. The competition of the channels of interaction of different symmetry and their orthogonality as $T \to 0$ leads also to logarithmic infra-red singularities  in the charge, spin and pairing channels. 
 
 While the results for the impurity have been obtained essentially exactly, it does not appear possible to extend similar methods to the  lattice. The best that one can hope for is a controlled  theory which in the limit of single impurity gives the important results of the exact solution, and compares well with the extra-ordinary experimental results in the lattice.
 
\begin{figure}
 \begin{center}
 \includegraphics[width= 1.0\columnwidth]{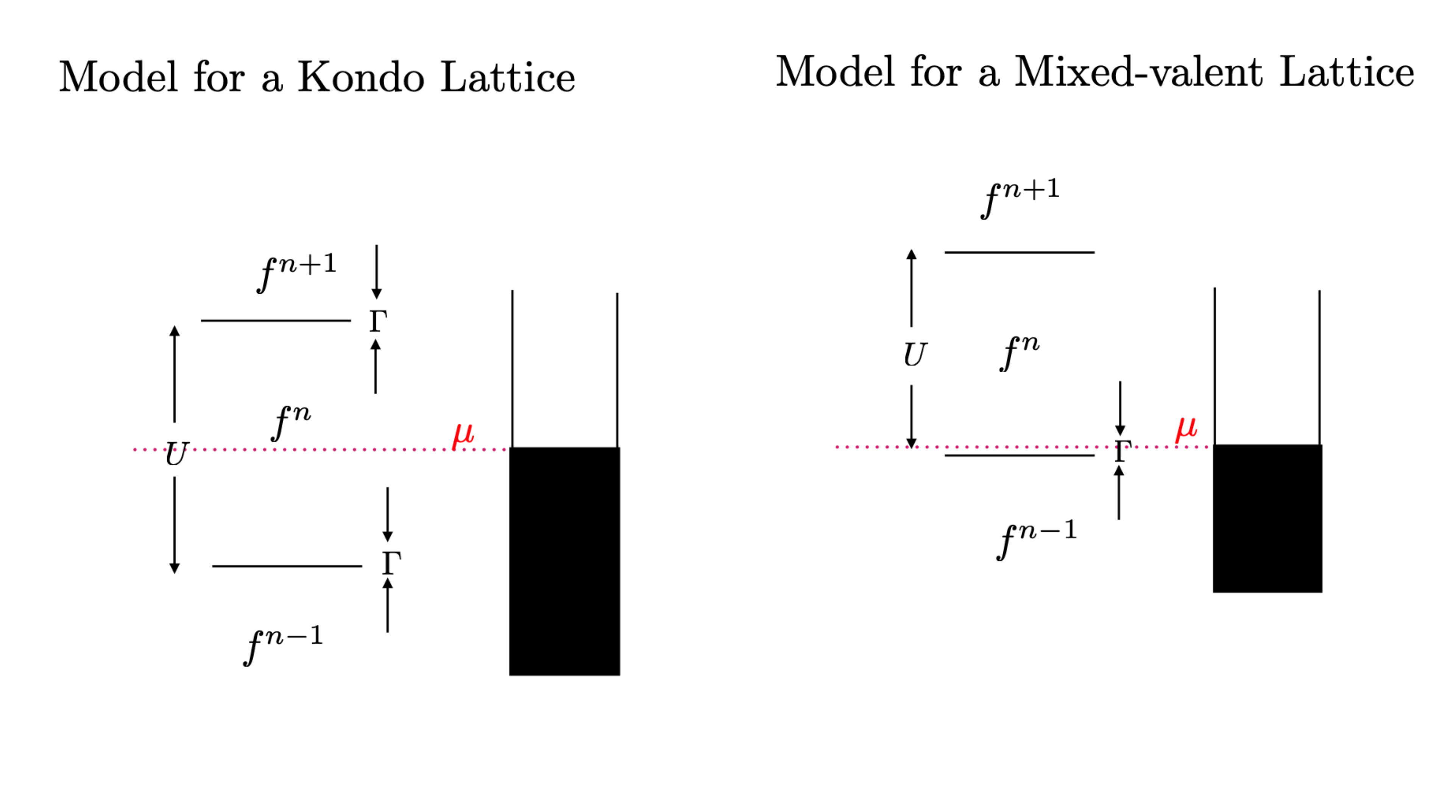}
 \end{center}
\caption{Difference of the model for the Kondo impurity in a metal from that for the  MIxed-valence impurity. The figures show the juxtaposition of the correlated f-levels for the two models with respect to the conduction band reservoir shown in black and the chemical potential, in terms of parameters defined in Eq. (\ref{H}). The dark horizontal lines give the location of the chemical potentials for an isolated correlated atom for a change of charge from one integral number to another. The necessity of obeying Friedel sum-rule or local charge neutrality both for the impurity charge state $n$ and the charge state $(n-1)$ requires at least one more interaction parameter than in the nearly particle-hole symmetric Anderson model for the Kondo effect. In the mixed-valence model, there is a competition between singularities for spin-fluctuations for a given charge state and a competition of charge fluctuations, leading to a quantum-critical point.}
 \label{Fig:KondoMV}
\end{figure}

The juxtaposition of the correlated f-levels with respect to the chemical potential in the band are sketched in Fig. (\ref{Fig:KondoMV}). 
The minimal model for the mixed-valence lattice on a basis of $f_{i \sigma}$ and $d_{i \sigma}$  orbitals at each site, has a Hamiltonian,  
\be
 \label{H} 
H &=&\sum_{i \sigma} H_{i \sigma} +  ~ \sum_{{\bf k}, \sigma} (\epsilon_d({\bf k}) -\mu) d^+_{{\bf k}, \sigma} d_{{\bf k} \sigma}, \\ \nonumber
H_{i \sigma} &=& \epsilon^0_f n_{f i}- \mu(n_{f i} + n_{di}) + \sum_n t_n  (f_{i\sigma}^+ d_{i+n, \sigma} + H.C.) + U n_{f i \uparrow} n_{f i \downarrow} \\ \nonumber
&+& (V-J)(n_{fi}-1/2) (n_{di}-1/2) + J(n_{fi \uparrow} n_{d,i \downarrow} + n_{di \uparrow} n_{f,i \downarrow}).
\ee
$n$ sums over the neighbors of a site i. 
 %The centroid $\epsilon_{d0}$ of the d-band energies $\epsilon(k)$  is at $0$.  
The correlated f-orbital energy is $\epsilon^0_f$ and $U >> t_n$.  On any given site f and d have different symmetry so that  neighbors of $i$ on sites at mutual reflection have opposite signs of $t_n$. For the same reason, interactions between both relative spin-combinations in $n_{fi}$ and $n_{di}$ is allowed through the parameter $V$ and there is a Hund's rule coupling $J$ favoring same spin.  
In the single-impurity problem $\overline{V} \equiv (V-J)$ is tuned to satisfy the stability condition for a mixed valence \cite{PerakisVRuck, Sire_V_R_G}. 
 I have subtracted $1/2$'s in that term so that with chemical potential $\mu =\epsilon^0_f$, the state with $(n_{fi} =0, n_{di} =1)$ has the same local energy as the state with ($n_{fi} = 1, n_{di} =0)$, thus maintaining local charge neutrality or the Friedel screening condition. If the f-level's energy renormalizes due to interactions away from $\mu$ as we shall find below, $\overline{V}$ must be adjusted to preserve this condition, provided stability of the solution for variation about the condition is also shown.
 
The simple model used here is sufficient to derive the properties for $<n_f> =1/2$.
In actual materials, the stable value of the fractional $f-$valence is determined by various other details \cite{Varma_Heine1975, V_RMP_MV} such as the ionicity and the radii of the cations and the anions. 

The equations of motion in the problem for $U \to \infty$ are equivalent to dropping the term proportional to $U$ and replacing the bare width of the f-level $\Gamma \equiv \pi t_n^2 \nu$, where $\nu$ is the d-band density of states at the chemical potential by the operator,
\be
\label{Gamma}
\overline{\Gamma}_{\sigma}  = \Gamma_{\sigma} (1-n_{f i -\sigma})
\ee
This procedure was introduced in Ref. \cite{V_Yafet1, V_RMP_MV} for the Anderson impurity model \cite{Anderson1961} as well as for a lattice; its mean-field value, $\Gamma_{\sigma} (1- <n_{f -\sigma}>)$, was shown to be the expression commonly used for the Kondo temperature. Similar result has been obtained with additional sophistication,  and other approximations in a variety of different ways, \cite{Barnes, Coleman1987, ReadNewns} \cite{MillisLeeHF, RamaSur1982}, \cite{Rice_Ueda_Gutzwill, V_Weber_Randall, KotliarRuckenstein}. The operators $(1- n_{f_i -\sigma})$ are often called the slave bosons. We may define $\overline{t}_{\sigma}$ by $\overline{\Gamma}_{\sigma} = \pi \overline{t}_{\sigma}^2 \nu$, which is marginal \cite{HaldanePRL1978, Krishnamurthy1980} about the high temperature fixed point where it flows to $0$ and the susceptibility is the Curie law. The crossover temperature is the Kondo temperature which is of $O(\overline{\Gamma})$ for mixed-valence \cite{V_Yafet1, Haldane1977_mixedval}. As will be seen below, we need not dwell on the issue of the precise value of $\overline{t}$.

\section{The Ansatz}

The ansatz introduced, to be tested variationally, is that beside their hybridization with the parameter $\overline{t}$, there is also an
 equal spin-odd parity  pairing of the $f$ and $d$ particles due to the local interaction proportional to $\overline{V}$ with center of mass momentum $2{\bf k}_F$ for all points ${\bf k}_F$ on the Fermi-surface. This is derived using the reduced Hamiltonian at each $2{\bf k}_F$, rather than  at  center of mass momentum $0$ as in the BCS reduced Hamiltonian. The total number of states is kept constant by appropriate normalization using $S_F$, the area of the Fermi-surface. The motivation to do this will be clear in the development below.  The particle and hole energy of the conduction band measured from the chemical potential are related for a range in their linear range of variation with ${\bf q}$, which is at least of 
 $O(\frac{\overline{t}}{W} k_F)$, where  $W$ is the d-electron bandwidth, so that
 \be
\label{qpe}
\epsilon_d({\bf k}_F + {\bf q}) -\mu = - (\epsilon_d({\bf k}_F - {\bf q})-\mu).
\ee
Using Eq. (\ref{qpe}), the Hamiltonian for the kinetic energy of the d-electrons may be written in terms of operators
 \be
 \label{basis}
\alpha_{d1}( {\bf k}_F, {\bf q}, \sigma) & \equiv &  \frac{1}{\sqrt{2}}(d({\bf k}_F+ {\bf q}, \sigma) + d^+({\bf k}_F-{\bf q}, \sigma)), \\
 \alpha_{d2}( {\bf k}_F, {\bf q}, \sigma) & \equiv &  \frac{i}{\sqrt{2}}(d({\bf k}_F+ {\bf q},\sigma) - d^+({\bf k}_F-{\bf q}, \sigma)),
\ee
as
\be
\label{choice}
\frac{1}{S_F} \sum_{{\bf k}_F,{\bf q}, \sigma, i=1,2} \big(\epsilon_d ({\bf k}_F + {\bf q}) -\mu\big) \alpha^+_{di}( {\bf k}_F, {\bf q}, \sigma)\alpha_{di}( {\bf k}_F, {\bf q}, \sigma).
\ee
In the usual choice, with the center of momentum as the $0$ of a Brillouin zone with center of symmetry, the kinetic energy is off-diagonal in the $\alpha$'s,
\be
\sum_{{\bf k}, \sigma} \big(\epsilon_d ({\bf k}) -\mu\big) \alpha^+_{d1}( {\bf k}_F, {\bf q}, \sigma)\alpha_{d2}( {\bf k}_F, {\bf q}, \sigma) + H.C.
\ee
The choice (\ref{choice}) will be quite important below.
The equal spin reduced pairing  Hamiltonian is 
\be
H_{pair} = \sum_{{\bf k}_F, {\bf q}, {\bf q}', \sigma} \overline{V} d^+({\bf k}_F- {\bf q},\sigma) f^+({\bf k}_F+{\bf q},\sigma) f({\bf k}_F + {\bf q}',\sigma) d({\bf k}_F- {\bf q}',\sigma).
 \ee 
The pairing amplitudes with center of mass momentum $2{\bf k}_F$ is introduced as
 \be
 P(2{\bf k}_F) =  \overline{V}(2 {\bf k}_F) \sum_{\bf q} <f({\bf k}_F - {\bf q},\sigma) d({\bf k}_F + {\bf q},\sigma)>; ~~P(2{\bf k}_F)= -P(-2{\bf  k}_F).
 \ee
Since this is a crucial step, it is worth  reiterating: For a translational invariant interaction Hamiltonian with interactions $V({\bf k, q, q'})$,  where ${\bf k}$ the center of mass momenta of a pair of holes or particles,  the BCS reduced Hamiltonian picks states with ${\bf k} =0$ for the  center of mass of a pair of holes or particles. I have picked the center of mass momentum to be ${\bf k} = 2{\bf k}_F$ for each ${\bf k}_F$ on the Fermi-surface and summed over the Fermi-surface. Usually such a pairing is disfavored because of the kinetic energy at such finite momentums. This is not an issue here because one of the partners in pairing are correlated $f$-electrons with effectively infinite mass.

 The hybridization Hamiltonian between the $d$ and the $f$-orbitals is 
\be
 \sum_ {{\bf k}_{F},{\bf q}, \sigma} t({\bf k}_F +{\bf q})  d^+({\bf k}_F + {\bf q},\sigma)f({\bf k}_F +{\bf q}, \sigma).
\ee
The odd-parity of the hybridization requires  $\overline{t}({\bf k}_F + {\bf q}) =  -\overline{t}(-({\bf k}_F + {\bf q}))$,
the same for both $\sigma$, just as for $P$.
 The effective Hamiltonian in the basis 
 $$\big( d ({\bf k}_F+ {\bf q}),~d^+ ({\bf k}_F- {\bf q}),~f({\bf k}_F+ {\bf q}),~f^+({\bf k}_F - {\bf q}) \big)^T$$
 \be
 \label{H1}
  H_{eff} =
  \frac{1}{S_F} \sum_{{\bf k}_F,{\bf q},\sigma} 
  \left(\begin{array}{cccc} \epsilon_d({\bf k}_F+ {\bf q})-\mu  &  0 & t({\bf k}_F) & P(2{\bf k}_F)\\0 & ( \epsilon_d({\bf k}_F + {\bf q})-\mu) & P(2{\bf k}_F) & t({\bf k}_F)  \\  t({\bf k}_F) & P(2{\bf k}_F) &  (\epsilon_f -\mu)& 0  \\  P^*(2{\bf k}_F) & t({\bf k}_F) & 0  & -(\epsilon_f- \mu)  \end{array}\right).
\ee
%\be
% \label{H1}
 %H_{eff} &=& \sum_{{\bf k}_F,{\bf q},\sigma} ({\epsilon}_d ({\bf k}_F+ {\bf q}) - \mu)d^+({\bf k}_F + {\bf q}, \sigma)d ({\bf k}_F+ {\bf q}) + ({\epsilon}_f  - \mu) %f^+({\bf k}_F+ {\bf q}, \sigma)f({\bf k}_F+ {\bf q},\sigma)  \nonumber \\
%&+& \overline{t}({\bf k}_F+ {\bf q})d^+({\bf k}_F+ {\bf q},\sigma)f({\bf k}_F+ {\bf q},\sigma) + \overline{V}~ P(2{\bf k}_F) ~ f^+({\bf k}_F- {\bf q}\sigma) d^+%%({\bf k}_F+ {\bf q}\sigma) + H.C.
 %\ee
 (\ref{H1}) is easily diagonalized but the Hamiltonian is more transparent in the basis of Eqs. (\ref{basis}) and their Hermitian conjugates and similarly defined $\alpha_1(f ({\bf k}_F+ {\bf q}, \sigma), \alpha_2(f  ({\bf k}_F+ {\bf q},\sigma)$ and their Hermitian conjugates. The $\alpha$'s have fermion commutations. 
Due to (\ref{qpe}), there is a symmetry under simultaneous charge conjugation and inversion about the Fermi-vectors for these fermions:
\be
\alpha_{d1,d2}({\bf k}_F, {\bf q}, \sigma) = \alpha^+_{d1,d2}({\bf k}_F, - {\bf q}, \sigma).
\ee
So also for the $\alpha_{f1,f2}({\bf k}_F, {\bf q}, \sigma)$.

\noindent
In terms of these operators, the mean-field effective Hamiltonian (\ref{H1}) 
%\be
%H_{eff} &=& \frac{1}{2} \sum_{{\bf k}_F, {\bf q}, \sigma}  (\epsilon({\bf k}_F+ {\bf q})-\mu) \big(\alpha^+(d 1 ({\bf k}_F +{\bf q}) \sigma) \alpha(d 1 ({\bf k}_F +{\bf q}) \sigma) + \alpha^+(d 2 ({\bf k}_F +{\bf q}) \sigma})\alpha(d 2({\bf k}_F + {\bf q}) \sigma)\big)  \nonumber \\
%&+& + i  (\epsilon_f-\mu)\alpha_1^+(f ({\bf k}_F +{\bf q}) \sigma)\alpha_2(f {({\bf k}_F +{\bf q})  \sigma)  \nonumber \\
%&+& i  (P(2{\bf  k}_F) + \overline{t}({\bf k}_F)) \alpha_2^+(f2 ({\bf k}_F +{\bf q})  \sigma) \alpha_1(d ({\bf k}_F +{\bf q}) \sigma) + i  (P(2{\bf k}_F) - \overline{t}({\bf k}_F)) \alpha_1^+(f ({\bf k}_F +{\bf q})  \sigma) \alpha_1(d  ({\bf k}_F +{\bf q})  \sigma).
% \ee
 written in the basis 
 \be
 \label{basis}
 \big(\alpha_{d1}({\bf k}_F, {\bf q}, \sigma), \alpha_{d2}({\bf k}_F, {\bf q}), \sigma) , \alpha_{f1}({\bf k}_F, {\bf q}, \sigma) ,\alpha_{f2}({\bf k}_F, {\bf q}, \sigma) \big)^T, 
 \ee
 identical for both components of spin, is 
   \be
   \label{Heff}
& & H_{eff} = \frac{1}{S_F} \sum_ {{\bf k}_F, {\bf q}, \sigma}  \nonumber \\
& & \left(\begin{array}{cccc}\epsilon_d({\bf k}_F+ {\bf q})-\mu  & 0 & 0 &i (P(2{\bf k}_F)+ t({\bf k}_F))\\ 0 &  \epsilon_d({\bf k}_F+ {\bf q})-\mu  & i (P(2{\bf k}_F)- t({\bf k}_F) & 0 \\ 0 & -  i (P(2{\bf k}_F)- t({\bf k}_F)) & 0 & i(\epsilon_f -\mu) \\ - i P(2{\bf k}_F)+ t({\bf k}_F)) & 0  & -i(\epsilon_f- \mu) & 0 \end{array}\right)
\ee
It is important that the first two terms in the diagonal are equal, in contrast to the case that the center of mass for pairing is zero for which the trace of the Hamiltonian matrix is zero. 

  As shown below, stability of the mixed-valence solution requires that the renormalized $f$ level be at the chemical potential, i.e. $\epsilon_f = \mu$. With this condition the eigenvalues of (\ref{Heff}) are
\be
\label{EVal}
\mathcal{E}_{(v,c) (1,2)}({\bf k}_F+ {\bf q}) = \frac{\epsilon({\bf k}_F+ {\bf q}) - \mu)}{2} \mp \sqrt{\Big(\frac{\epsilon({\bf k}_F+ {\bf q}) - \mu}{2}\Big)^2+ |\overline{t}({\bf k}_F) \mp P(2{\bf k}_F)|^2}.
\ee
Since variation of  $t({\bf k}_F +{\bf q})$ with ${\bf q}$ is negligible compared to the variations in $\epsilon_d({\bf \bf k}_F +{\bf q})$, it has been ignored.

\section{ Determining $P(2{\bf k}_F)$}
\subsection{Self-consistency condition}
 The ground state energy calculated from  $<H_{eff}>$ is minimized with respect to $P(2{\bf k}_F)$  for a fixed $\overline{t}({\bf k}_F)$ and  with a lower-cutoff $- W$ for the d-band. This gives the relation between $P(2{\bf k}_F)$ and $\overline{t}({\bf k}_F)$, (unless explicitly required, $P(2{\bf k}_F) \equiv P,  \overline{t}({\bf k}_F) \equiv \overline{t} $ and the partial density of states  at the chemical potential $\nu({\bf k}_F) = \nu$ below),
\be
P = \overline{V} \int_0^{-W} d\epsilon~ \nu \Big(\frac{(P + \overline{t})}{\sqrt{\epsilon^2/4 + (P+ \overline{t})^2}} + \frac{(P- \overline{t})}{\sqrt{\epsilon^2/4 + (P -\overline{t})^2}}\Big).
\ee
where  $\lambda \equiv \lambda({\bf k}_F) \equiv \nu({\bf k}_F) \overline{V}$, this gives in terms of  $r \equiv r({\bf k}_F) \equiv P(2{\bf k}_F)/{\overline{t}}({\bf k}_F)$,
\be
\label{cond}
\Big(\frac{1}{\lambda} - 2\ln \frac{W}{\overline{t}}\Big) = \frac{1}{r}\Big((r+1)\log |r+1| + (r-1) \log |r-1| 
\Big).
\ee
$r = 0$ is one obvious solution. On examining the curvature, the solution $r =0$ has a local minimum only for $\lambda < \frac{1}{2}(1+ \ln(W/{\overline{t}}))^{-1}$. The other solutions are symmetric about $r=0$ and close to $\pm 1$. We can confine further discussion for $r \ne 0$ to its positive values. 

\subsection{Condition on $r$ from stability of mixed-valence.}

Eq. (\ref{cond}) must be supplemented by the stability condition for mixed-valence, that the average occupation of the $f$ and the $d$ states is stable for each at $1/2$ per site.  This is satisfied at the Hartree-Fock level on taking the expectation values $n_{fi} \to <n_f>, n_{di} \to <n_d>$ by the starting Hamiltonian, Eq. (\ref{H}). We must now consider the fluctuation contribution. To that end, the leading term in the self-energy $\Sigma_f(i,\omega)$ of the f-level is calculated. This self-energy may be written using the eigenvectors and eigenvalues of (\ref{H1}),
\be
 \label{SEf}
 Re ~\Sigma_{fi}(\omega) &=& \frac{1}{S_F} \sum_{{\bf k}_F, {\bf q},\sigma} |P^2(2{\bf k}_F) - \overline{t}^2({\bf k}_F)| ~  < d^+_{{\bf k_F +q},\sigma}d_{{\bf k_F+q},\sigma}>  \nonumber  \\
 &=& -  \frac{1}{S_F} \sum_{{\bf k}_F} \nu_{{\bf k}_F}  |P^2(2{{\bf k}}_F) - \overline{t}^2({{\bf k}}_F)| ~\log\big| \frac{W}{\omega} \big|.  
 \ee
The renormalized $f$-level is given self-consistently by $\epsilon_{f,i} -\mu = \epsilon^0_{f,i} - \mu - Re ~\Sigma_{fi}(\epsilon_{f,i} -\mu)$.  For  $(\epsilon_f  -\mu) << \nu({\bf k}_F) |P^2 - \overline{t}^2|$, the solution is 
\be
(\epsilon_f  -\mu) = - c ~ \frac{1}{S_F} \sum_{{\bf k}_F} \nu({\bf k}_F) |P(2{\bf k}_F)^2 - \overline{t}({\bf k}_F)^2| . 
\ee
 $c$ is a numerical factor of $O(1)$. For the eigenvectors corresponding to Eq. (\ref{EVal}), the stable insulating state requires the chemical potential to be at the mid-point of the minimal gap between $\mathcal{E}_{(v,1)}({k})$ and $\mathcal{E}_{(c,1)}({k})$ so that the particle and hole excitation energies are equal; only then is the compressibility well defined for a pure insulator. This happens only for $\epsilon_f  = \mu$. So we require $ |P(2{\bf k}_F)^2 - \overline{t}({\bf k}_F)^2| = 0 $, or $r_{{\bf k}_F} = \pm 1$, i.e. at the non-analyticity of Eq. (\ref{cond}). The higher order corrections to this self-consistent non-crossing approximation, also vanish only for all $r  =1$.

We now look for a simultaneous solution of this condition for stability of mixed-valence with the other equation for $r$, i.e. Eq. (\ref{cond}). We then require satisfying the two equations
 \be
 \label{cond2}
 \frac{1}{\lambda} - 2 \log \frac{W}{2\overline{t}} &=& 0, \\ 
  \frac{r-1}{r} \log |r-1| &=& 0.
\ee
The former is the condition on $\lambda = \nu \overline{V}$ to retain the mixed-valence condition $<n_f> = <n_d> =1/2$ per site, and an incompressible insulating state, which was described as the necessary condition on the parameters of  the model Hamiltonian, (\ref{H}). The latter also has a solution at $r = 1$. It should be appreciated that the physical conditions for mixed-valence at values anywhere $0 < <n_f> <1$ may exist in any given situation. Without introducing any additional parameters, this may be reproduced in the model by the appropriate numbers subtracted from the operator $n_{fi}$ and  $n_{di}$ in the term in the starting model Hamiltonian, Eq. (\ref{H}), with coefficent $(V-J)$.

  We must examine also that the system is stable about this solution. 
 We find by calculating  the  second derivative of the energy with respect to $r$, that it is positive for $r = 1$ in the sizable range $O(\pm W/4)$. 

The condition on $P(T)$ at finite temperatures from minimizing the free-energy is
\be
\frac{P}{\overline{V}} = (P + \overline{t}) \int^{-W}_{E_{0+}} dE_+  ~\nu(E_+) \frac{\tanh(\beta E_+/2)}{\sqrt{\epsilon^2 + (P+{\overline{t}})^2}}
+(\overline{t} \to -\overline{t}, E_+\to E_-)
%P-\overline{t})\int_{E_{0-}}^{-W} dE_-~\nu(E_{-}) \frac{\tanh(\beta E_-/2)}{\sqrt{\epsilon^2 + (P - {\overline{t}})^2}}.
\ee
 $E_{+,-}$ refers to the eigenvalues in Eq. (\ref{EVal}) with $(\overline{t} \pm P)$. $E_{0,+,-}$ are the lowest values $|\overline{t} \pm P|^2/W$ allowed of $E_{+,-}$ and $\nu(E) \equiv \nu(\epsilon) \frac{d \epsilon}{d E}$.
Noting that
$(\epsilon^2 + (\overline{t} \pm P|^2)^{-1/2} \frac{d \epsilon}{dE_{\pm}} = E^{-1}_{\pm}$, together
with
$\int_a^W dx \frac{\tanh(\beta x)}{x} = \ln (W/a)$
for $\beta W >> 1$, gives the condition $r(T) = \pm 1$ or $0$ independent of temperature with the same conditions as at $T=0$. Note that ${\overline{t}}$ itself crosses over to $0$ as the high temperature fixed point is approached. 
There are also analytic finite temperature corrections about $T=0$ as for ordinary insulating and metallic states
for the two pairs of sets of states. 

%\section{ The Principal Physical Results}

\subsection{Eigenvalues and eigenvectors for single-particle excitations}

\begin{figure}[b]
 \begin{center}
 \includegraphics[width= 1.0\columnwidth]{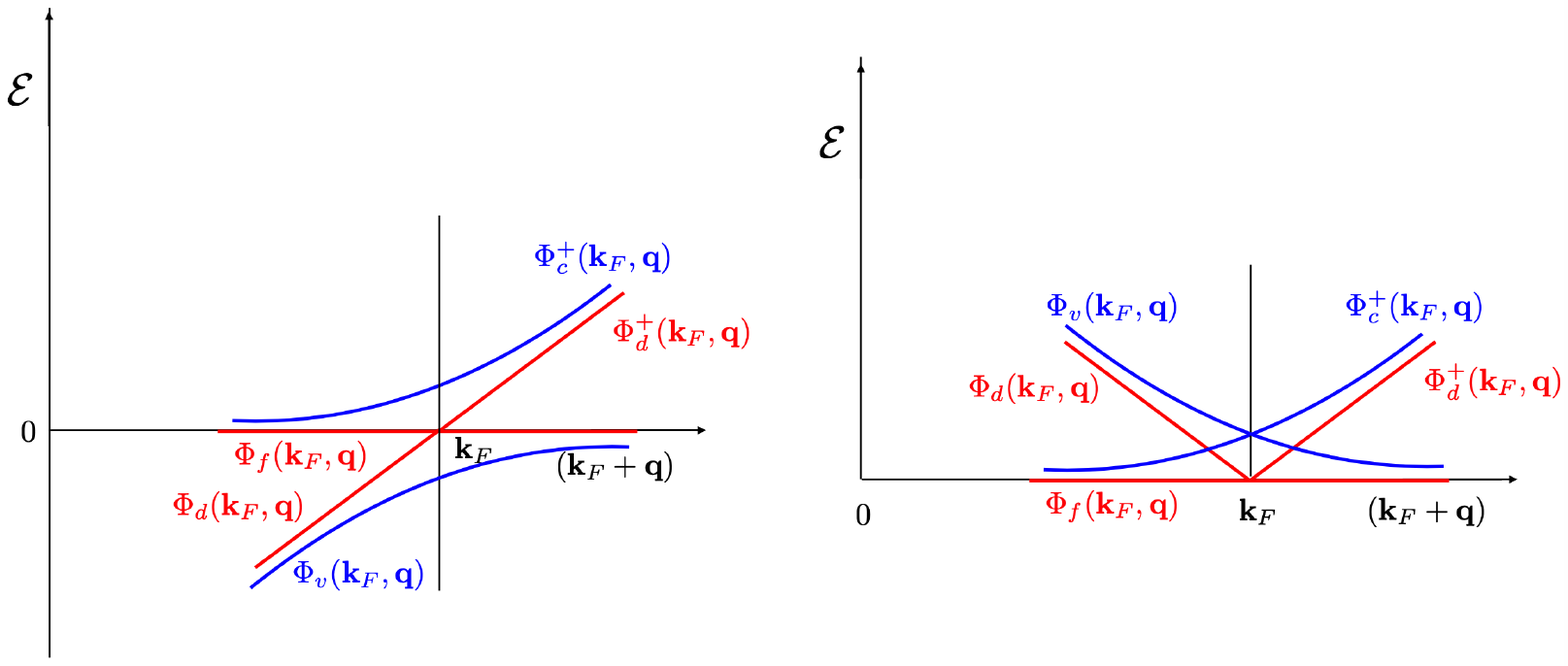}
 \end{center}
\caption{Single-particle spectra with wave-functions indicated. On the left it is shown as conventional drawing of band-structure and on the right as the excitation spectra. Both are shown only for the part where the bare d-band dispersion is linear with departure of momentum ${\bf q}$ from the bare d-band  Fermi-vectors ${\bf k}_F$.  The calculations give the fractionalized excitations $\Phi_f({\bf k}_f, {\bf q})$ and $\Phi_d({\bf k}_f, {\bf q})$ which are linear combinations of bare particles $({\bf q} > 0)$ and holes $({\bf q} < 0)$ for the ${\bf k}_F$ shown, identical for both spins, with dispersion identical to the bare dispersion of the d-band and the correlated f-levels, and $\Phi_v({\bf k}_f, {\bf q})$ and $\Phi_c({\bf k}_f, {\bf q})$ which are hybrids of $d$ and $f$ states.}
 \label{Fig:SinglePartSpectra}
\end{figure}

For $P = \overline{t}$, the eigenvalues and eigenvectors may be read off from (\ref{Heff}). The eigenvectors are normalized so that the total number of fermions is conserved. Two of the eigenvalues at each momenta and spin are unaffected by interactions or hybridization and are at $(\epsilon_d({\bf k}_F+{\bf q}) -\mu)$ and $(\epsilon_f-\mu) = 0$. Their respective eigenvectors are 
\be
\label{majo1}
\Phi_{d}({\bf k}_F, {\bf q}, \sigma) &=&   \frac{i}{{2}}(d({\bf k}_F+ {\bf q}, \sigma) - d^+({\bf k}_F-{\bf q}, \sigma)), \\
\label{majo2}
\Phi_{f}({\bf k}_F, {\bf q}, \sigma) &=&   \frac{1}{{2}}(f({\bf k}_F+ {\bf q}, \sigma) + f^+({\bf k}_F-{\bf q}, \sigma)),.
\ee
Since,
\be
\Phi_{d}({\bf k}_F, {\bf q}, \sigma) &=& \frac{1}{2} \sum_i \big((d_{i,\sigma} + d^+_{i,\sigma}) \cos({\bf k}_F.{\bf R}_i) + i (d_{i,\sigma} - d^+_{i,\sigma}) \sin({\bf k}_F.{\bf R}_i)\big) e^{i{\bf q}.{\bf R_i}},    \\
&=& \Phi^+_{d}({\bf k}_F, - {\bf q}, \sigma), ~and \\
\Phi_{f}({\bf k}_F, {\bf q}, \sigma) &= & \Phi^+_{f}({\bf k}_F, -{\bf q}, \sigma),
\ee
these fermions are real or Majoranas at ${\bf q} =0$, i.e. at all points at the Fermi-surface. But being equal linear combinations of ordinary particle and holes, they carry no charge at any $({\bf k}_F, {\bf q})$. Being also equal linear combinations of such particle and holes with the same spin, they also carry no magnetic moment, but they retain their doublet degeneracy. This point will arise again in a discussion of  the magnetization in an applied field. 

The other two eigenvalues are hybridization of the d-band with the localized f-levels, 
\be
\label{ev}
E(v,c) = \frac{1}{2}\big((\epsilon_d({\bf k}_F+{\bf q}) -\mu) \pm \sqrt{(\epsilon_d({\bf k}_F+{\bf q}) -\mu)^2 + 16 |t({\bf k}_F)|^2}\big).
\ee
The eigenvectors for the hybrid states can be specified in terms of an angle $\theta({\bf k})$,
\be
\tan \frac{\theta({\bf k})}{2} \equiv \frac{4\overline{t}}{\epsilon_d({\bf k})}, ~u_{v,c}({\bf k}) \equiv \frac{\sin \frac{\theta({\bf k})}{2}}{\sqrt{1 \mp \cos
 \frac{\theta({\bf k})}{2}}}
\ee
Given their spectra, they may be thought of as  valence band and conduction band states. On the basis of (\ref{basis}, they are
\begin{equation}
\label{v}
\Phi_{v}({\bf k}_F, {\bf q}, \sigma) = \frac{1}{{2}} \Big( 0, ~i u_v({\bf k}), ~ 0, ~ - i \sqrt{1- u^2_v({\bf k})}\Big), 
\end{equation}
\begin{equation}
\label{c}
\Phi_{c}({\bf k}_F, {\bf q}, \sigma) = \frac{1}{{2}} \Big( 0, ~i u_c({\bf k}), ~ 0, ~ - i \sqrt{1- u^2_c({\bf k})}\Big).
\end{equation}

\section{ The Principal Physical Results}

\subsection {Magneto-oscillations}
  
  Let us consider the change in the Hamiltonian, Eq. (\ref{Heff}) due to the orbital effects of a magnetic field ${\bf H}$ for the cyclotron energy much less than the gap  $4|(\overline{t} +P)|$. As discussed below and elsewhere \cite{CMVcorrins}, for larger magnetic fields, there should be a transition due to Zeeman effect from a mixed-valence insulator to a metal. ${\overline{t}}$  is a  nearest neighbor hybridization parameter which is unchanged by orbital effects of the magnetic field. The self-consistent value of $P$ will change due
  to orbital effects on the kinetic energy. This effect is negligible until the cyclotron energy is comparable to the zero field value of $P$.  Let us stay well below such a value so that the condition $P = {\overline{t}}$ due to mixed valence and energy minimization is retained.
  
  The only changes to consider are in the diagonal terms in Eq. (\ref{Heff}), which require changing their momentum dependence to that of $({\bf k}_F +{\bf q} + \frac{e}{c} {\bf A})$, where ${\bf A}$ is the vector potential due to the applied magnetic field. The first diagonal term is hybridized and has a gap. Therefore we need  consider only the second diagonal term which is free with eigenvalues in zero field,  $(\epsilon_d({\bf k}_F+{\bf q}) -\mu)$, which  are the same as the bare unhybridized d-band-structure. The effective Hamiltonian for such states in Eq. (\ref{Heff}) in a field commutes with the rest of the Hamiltonian under the conditions stated and may be diagonalized to form Landau level spectrum as usual. Since the number of particles in this state is also separately conserved, the harmonic oscillator quantum number in a magnetic field crosses the chemical potential as the cyclotron energy is varied and the flux quantization condition remains as in the ordinary case. Magneto-oscillations with $1/H$ periodicity therefore follow. 
  
  Given the normalization of the states, the amplitude of the oscillations should be 1/2 that is customary as also the Landau diamagnetism from these states.  So completely normal deHaas van-Alphen and other oscillations are to be expected except that the amplitude of the oscillations will be $\frac{1}{2}$ the normal value. This prediction can be tested by comparing with the amplitude on further increasing the field, where as discussed below, the "insulator" turns to a metal. The temperature dependence of their amplitudes are determined by the Fermi-distribution just as in the Lifshitz-Kosevich calculations. 
 
%A related interesting problem, which we do not derive here, is the relation in second quantization of the eigenvectors in a magnetic field to Eq. (\ref{majo1}) without a field. 
  
   Given their very large effective mass, the correlated $f$ orbital wave-functions are unaffected orbitally with or without $P$ by the magnetic field. 
  
  Let us now consider the Zeeman effect on the magnetic oscillations. 
   \be
H_Z = - g \mu_B \sigma \cdot H \sum_i  \big( g_f f_{i\sigma}^+f_{i\sigma} + g_d d_{i\sigma}^+d_{i\sigma}\big).
\ee
Consider the effects perturbatively on the states (\ref{majo1})  crossing the chemical potential. It is easy to check that the Zeeman magnetic susceptibility of these bands is $0$ because, integrated over $k$, they have equal linear combinations of hole and electron states with the same spin. The Fermi-wave-vector for the up and down spin remains the same. There can then be no spin-splitting of the magneto-oscillations. This is an important prediction.

  \subsection{Thermodynamics}
   \subsubsection{Specific heat and Entropy}
  At temperatures small compared to the gap, one needs consider the specific heat only due to the excitation of the states (\ref{majo1}) and (\ref{majo2}). The former of-course lead to a free-energy identical to any other fermions and therefore a specific heat linear in temperature with a magnitude given by the area of the fermi-surface and the effective mass. The agreement of the measured specific heat from such parameters determined by the magneto-oscillations in $YbB_{12}$ is one of the most conclusive experimental results for propagating fermions which is electrically an insulator \cite{Sato2019}. The agreement of the measured thermal conductivity \cite{Sato2019} with the deduced fermion velocity and an estimated density of impurity scattering from other measurements further confirms the existence of degenerate fermions consistent with the magneto-oscillations.
    
  $SmB_6$ also has a large specific heat at low temperatures \cite{SebastianSmB62020, PhelanSphtSmB6}, larger than what one would expect from the magneto-oscillation experiments, but its linearity with temperature is unclear.  It may also  have contribution from impurity states in the gap. Samples prepared differently \cite{SebastianSmB62020, PhelanSphtSmB6} give significantly different low temperature specific heat. The case for the linear in T specific heat cannot be made as convincingly as in YbB$_{12}$. Moreover, below about $1 K$, it shows a very rapid increase indicative of a transition or a cross-over to some other state.
  Not coincidentally $\mu$-sR relaxation rate measurements \cite{BiswasMusR} show a constant density of magnetic excitations below temperatures where the specific heat shows its giant rise.
  
  The states (\ref{majo2}) present interesting issues. Being localized correlates states (at the chemical potential),  we must consider the entropy of their spin-degrees of freedom even though they are dark at low energy to electromagnetic  perturbations. They  have a free-energy independent of temperature, so that they have ground state entropy. The magnitude of this is large of $O(R \ln 2)$. If residual interactions remove their entropy, they must lead to some exotic order. Indeed the very rapid increase in specific heat at low temperatures is indicative of new physical phenomena at low temperatures.  A possible way to test for remnant entropy at least up to very low temperatures is the difference in integrated $C_v/T$ from such temperatures as a function of magnetic fields up to temperatures much larger than $\mu_B H/k_B$. The prediction that quantum-mechanics allows extensive remnant entropy (without disorder or frustration) may be the most important new  result in this work.
  
\subsection{Response Functions} 

States $\Phi_v$ and $\Phi_c$ are admixtures of $f$ and $d$ states with an energy gap. These fermions have complex wave-functions except at some special momenta as given by the coefficients $u_c({\bf k}), u_v({\bf k})$. One expects transitions between them by electromagnetic fields as well as by neutron scattering. One also expect to see the $\Phi_v$ states in ARPES and both $\Phi_v$ and $\Phi_c$ states in tunneling measurements. Since the relative admixture is affected by the usual particle-hole coupling between the $f$ and the $d$ states as well as the particle-particle admixing, the "coherence-factors" are different from the normal semi-conductor. So  the amplitudes variations with momentum transfers at energies above the gap are quantitatively different from the simpler case of particle-hole mixing.

However, much more interesting are the response from the states which cross the chemical potential, i.e. the $\Phi_d$ and $\Phi_f$ states.

Let us consider the coupling to the d-electrons first. The response functions are calculated from external perturbations which may in general be specified by operators 
\be
\label{pert}
H_{dd}' = \sum_{{\bf k, k}', \sigma, \sigma'} \sum_{a,b = f,d} M_{ab}({\bf k, k'}, \sigma, \sigma') a^+_{{\bf k}', \sigma'}  b_{{\bf k}, \sigma} + H.C. 
\ee
which specifies the momentum transfer ${\bf k'-k}$ and spin-scattering $(\sigma \to \sigma')$. The external fields may also be at a various energies. $H'$ causes transitions if it has a matrix elements in the eigenstates $(\Phi^+_a({\bf k}_F, q, \sigma), (\Phi_b({\bf k'}_F, q', \sigma'))$. $d_{{\bf k}, \sigma}, f_{{\bf k}, \sigma}$ or their hermitian conjugates cannot be expressed in terms of linear combinations of these states at the same energy. Let us consider all their products and drop terms with two creation and annihilation operators for $d-$fermions to which $H'$ does not couple. One then gets two kinds of terms
\be
\label{pert2}
(i): & \Phi_d^+({\bf k'}_F, q', \sigma') \Phi_d ({\bf k}_F, q, \sigma) \to (d^+_{{\bf k'}_F + {\bf q'}, \sigma'} d_{{\bf k}_F + {\bf q}, \sigma}
 - d^+_{{\bf k'}_F - {\bf q'}, \sigma'} d_{{\bf k}_F - {\bf q}, \sigma}). \\ 
 (ii): & ~~\Phi^+_d({\bf k'}_F, -q', \sigma') \Phi_d( {\bf k}_F, q, \sigma) \to (d^+_{{\bf k'}_F - {\bf q'}, \sigma'} d_{{\bf k}_F + {\bf q}, \sigma}
 - d^+_{{\bf k}_F - {\bf q}, \sigma} d_{{\bf k}'_F + {\bf q}', \sigma'}). 
 \ee
 %Let us now consider various experiments which reveal the low energy particle-hole response functions.
 
 \subsubsection{Electromagnetic absorption}
  Let us first consider the low energy absorption including the dc conductivity.   The perturbation is proportional to 
  ${\bf j}. {\bf A}$, where ${\bf j}$ is the current operator and ${\bf A}$ is the vector potential. The matrix elements of the current operator between normal fermions at momentum ${\bf k}_F + {\bf q}$ and ${\bf k}_F + {\bf q}'$ give contribution $\propto 2{\bf k}_F + {\bf q +q'}$. After taking the matrix elements one puts ${\bf q} + {\bf q'} \to 0$ and calculates the frequency dependence of the current-current correlations for response at spatially nearly uniform fields. 
  Acting on $\Phi_{d,f}({\bf k}_F, {\bf  q})$, the current operator gives a term $\propto  {\bf q}$, because as noted above $\Phi_{d,f}({\bf k}_F, {\bf  q})$ are purely real for ${\bf q} \to 0$. The current operator acting on a real wave-function or a Majorana is obviously $0$. Then taking the long wave-length limit, the current-current correlation from gap-less states is $0$.  There can be no finite conductivity at zero or finite frequency optical experiments.
  
The same is true of the contribution to  the current-correlations between the gapless states and the hybridized states  hybridized states $\Phi_{c,f}$ which have a gap. 

On the other hand, the long-wavelength current correlations from transitions between $\Phi_v({\bf k}_F, {\bf  q})$
and $\Phi_c({\bf k}_F, {\bf  q})$ are not zero and their value is related to the density of states and the value of the coefficents $u_v, u_c$ at the ${\bf k}_F$'s.

 \subsubsection{Ultrasound attenuation:}
  Longitudinal ultrasound attenuation is measured by a spin-conserving perturbation with absolute magnitude for small momentum transfer $|{\bf p}|$ for longitudinal sound and small energy transfer. So we put $(\sigma = \sigma')' , {\bf k}_F = {\bf k'}_F$ in the d-states crossing the chemical potential It is easily seen that the matrix elements of $M$ cancel  between the two terms, both of (i) and of (ii) in (\ref{pert2}). There is no ultrasound attenuation due to coupling to fermions. Transverse ultrasound attenuation is related ultimately to electromagentic absorption and so there is none either.

  \subsubsection{NMR:}
For NMR, one must consider transitions in the limit of the very low NMR frequency and  both $\sigma \ne \sigma'$ and ${\bf k}_F \ne {\bf k'}_F$. For the states crossing the chemical potential, the matrix elements are again $0$ because they are equal admixtures of holes and particles with the same spin. So there is no Korringa nuclear relaxation characteristic of metals is possible.
  
  \subsection{ The Single-Particle Spectra:}  

 %Eqs. (\ref{majo1}) give a "free" fermion excitation spectra with annihilation operators $\Phi_{d}({\bf k}_F, {\bf q}, \sigma)$ and non-dispersive fermions with annihilation operators $\Phi_f{({\bf k}_F, {\bf q}, \sigma)}$  without a gap at the chemical potential. The former are composed of pure d-states and the latter of pure f-states. These excitations are  shown in red in Fig. (\ref{wavefns.pdf}). The other two are gapped valence and conduction band states with eigenvectors  which are linear combinations of $f$  and $d-states$, but as evident from Eqs. (\ref{v, c}), far from ${\bf k}_F$, they turn to pure $d$ and $f$ states. Their energies are then very close $O(\overline{t}^2/W)$ of the free states at such points. If their energy difference is less than their thermal or natural linewidths, they may be  combined to form $d$ and $f$ ordinary fermion states. Very close to ${\bf k}_F$, where $\theta/2 \approx  \pi/4$, the states are linear admixtures of $f$ and $d$ states. 
 
 In photoemission (PE) experiments or Angle-resolved photoemission (ARPES), the external probe couple to the single-particle excitations through a the current operator acting on a bare annihilation operator and momentum and high energy fermions which may be considered ordinary free-fermions. The initial electronic states are at low energy and near ${\bf k}_F$. So for the same reason as for optical conductivity, they will remain dark in such experiments. 
 %There should then be an evolution of low energy states discovered by  such experiments as the momentum of the incoming photons changes from the optical to the x-ray range. For a fixed low energy, the states $\Phi_{d}({\bf k}_F, {\bf q}, \sigma)$ and $\Phi_f{({\bf k}_F, {\bf q}, \sigma)}$ have a bare annihilation operator which has $1/2$, the usual amplitude. So a Fermi-edge in ARPES  of $1/4$, the normal spectral weight is to be expected. A Fermi-edge is indeed observed continuing to low temperatures in high resolution PE measurements \cite{YbB12_2015_HybGap}, which normalized to the excitations above the gap has a spectral weight consistent with this value.
% However, it is unclear whether such states are bulk states (as studied here) or surface states. PE with photons which separate the bulk contribution from the surface may be needed to settle the issue. 
 
 In planar tunneling and STM experiments, the Hamiltonian is
 \be
 \sum_{{\bf k,k}' ,\sigma}T_{\bf k,k'} c^+_{{\bf k}, \sigma} d_{{\bf k}', \sigma} +H.C.
 \ee
 Here $c, c^+$ refer to operators of the ordinary metal or metal tip. In these experiments, there is no reason for the matrix elements to be zero. The tunneling current to the low energy states is however reduced, depending on the mixed valence and will be $1/4$, the normal value for example calcualted in this paper.  Tunneling and STM experiments do indeed see excitations in the insulating gap both in SmB$_6$ and YbB$_12$. But again, there is no clear evidence for whether they are surface states or bulk excitations or both.

\subsection{Superfluid Response?}

Since pairing has been used in the formulation of the theory, it is natural to ask if there is any superfluid response.
It can be ruled out if there is conservation of particle-number, which leads to zero superfluid density or zero superfluid density due to  other reasons.

The gapless excitations, Eqs. (\ref{majo1}, \ref{majo2}) are made up of pure $d$ and $f$ states and do not invoke
$f-d$ pairing. From the Eqs. (\ref{majo1}) and (\ref{majo2}), it is easily shown that particle density fluctuation calculated for them: $<(\delta n_d)^2> = <(\delta n_f)^2> = 0.$

The gapped excitations, Eqs. (\ref{c}, \ref{v}) present a different story since the gap is determined, beside the hybridization  $\sum_n t_n(<f^+_{i, \sigma} d_{i,+n, \sigma}> + H.C.$ of the $f$ and the $d$ states, by the pairing between them, $<f_{i,\sigma}d_{i,\sigma}>, <d^+_{i,\sigma}f^+_{i,\sigma}>$. There are many ways one can see that the superfluid density is zero. We can borrow from  Nozi\`eres and Pistolesi \cite{Nozieres1999} who investigated pairing in band insulators due to effective interactions of the excitations. They show that there can be no superfluid response as long as the gap in the excitation spectrum is non-zero. The reason is that giving a small finite momentum ${\bf Q}$ for the center of mass motion of the ground state leads to states with no change in energy as long as there is a gap; one simply moves all the states by ${\bf Q}$ without change in energy. In the present case, there is the additional argument that the center of mass linearly depends on the bare $f$-electron mass which is $\infty$ due to correlations. Finally, one can use the calculations showing that in an insulator the diamagnetic current  response is exactly cancelled by the paramagnetic inter-band response. The last is really a  muscular way of implementing the Nozi\`eres-Pistolesi argument.

%We can also do the following: The amplitude $|P_i| = {\overline{t}}$ at every site has been found in the mean-field calculation above. We must allow $P_i = P e^{i\phi_i}$ and calculate the static fluctuations $<P_i^* P_j>$. The correlations can be derived from the fluctuation contribution of the term in $H$ proportional to $V$; the slowest decaying contribution is 
%\be
%<P_i^* P_j> = V^2 P^2 \sum_n  <d_i^+ d_j>(- \omega_n) <f_i^+ f_j>(\omega_n).
%\ee
%The single-particle Green's function $<d_i^+ d_j>(\omega_n)$ has projection on the fermion band $\alpha$ which is gapless at the chemical potential. However, $<f_i^+ f_j>(\omega_n)$ has projections only on the bands $v$ and $c$ with a gap $O({\overline{\Gamma}})$. With the velocity $\to 0$ in these bands over a large part of the Brillouin zone, $<P_i^* P_j>$ decays exponentially with a  length of no more than a lattice constant. This result means that the phase of $P_i$ is independent at every site. 

%There is no long-range condensation nor a phase transition as a function of temperature in this model.  $|P|$ follows ${\overline{t}}$ at any temperature and acquires a finite value through a cross-over in temperature. 
%It is shown below that the superfluid density is $0$, as in an insulator. 

\subsection{Magnetization at high fields}

 We have already considered the lack of spin-splitting in the states (\ref{majo1}, \ref{majo2}) at small fields.
The gapped bands  $\Phi_v({{\bf k}, \sigma})$ and $\Phi_c({{\bf k}, \sigma})$ are also similar combination of spins for particles and holes as Eqs. (\ref{majo1}, \ref{majo2}). So their  susceptibility at low field  is also zero.  The finite magnetic susceptibility in the experiments has the magnitude of a van-Vleck susceptibility due to virtual transitions between the $f$-part of the states separated by a gap, which depend also on  the magnitude of the spin-orbit splitting of the f-orbitals. Such details are absent in the model above.  

The perturbative calculation of the response to a static magnetic field breaks down as  $g\mu_B H$ is increased to the order of the gap $\overline{\Gamma}$. Now we must re-consider the relation  Eq. (\ref{Gamma}) between $\overline{\Gamma}_{\sigma}$ and $<1- n_{f,-\sigma}>$, which marks the end of the Kondo or mixed-valence renormalizations. As argued earlier \cite{CMVcorrins}, the occupation of the minority $f$-spin must go to $0$ and the occupation of the majority $f-$spin $\to 1/2$, maintaining the occupation required at mixed-valence by the condition that $\overline{V} >> t$. This marks a transition \cite{CMVcorrins} from the insulator to a (polarized) metal where the effective ${\overline{t}}_{\downarrow} \to 0$ so that the hybridization of the wide band with this spin disappears.  Near this point, the self-consistency condition continues to give  $P_{\sigma} = \pm \overline{t}_{\sigma}$ on minimizing the energy with respect to $P_{\sigma}$ for a fixed $\overline{t}_{\sigma}$. 
 Then we can write $(\overline{t}_{\sigma} + P_{\sigma}) = \zeta (H-H_c)$, where $H_c$ is the field for the transition to the metallic state, so that the dispersion of band at the chemical potential of one of the spins has $(\zeta (H_0-H_c))^2$ replacing $(\overline{t}+P)^2$. The {\it increase} of magnetization for $H_0 >H_c$ then starts with
\be
\label{magn}
M \propto  \sqrt{2H_c(H_0-H_c)}.
\ee
 A sharp change in the magnetization just above the insulator to metal transition in a field was reported \cite{Singleton2021, Singleton_aps_2023} and appears to be consistent with Eq. (\ref{magn}). Also, no Zeeman splitting of oscillations was observed in the insulating state while it appears in the metallic state \cite{Singleton2021, Singleton_aps_2023}. These results are a strong tests of  the theory given here. Interestingly, the torque in the experiments increases dramatically above $H_c$. I attribute this to the re-appearance of spin and therefore of spin-orbit coupling for the low energy states in the ordinary metallic state arrived at in a large enough field.

\section{Concluding Remarks and Summary}

The results in this paper are based on two physical considerations. First is the obvious, but unused, fact that in a conduction band the energy of excitations $\epsilon({k}_F + {\bf q}) -\mu = - (\epsilon({k}_F - {\bf q}) -\mu)$. This  leads in any dimension to fermions with small departure from the Fermi-surface ${\bf q} =0$ that are a small departures from the pair of orthogonal Majorana excitations precisely at the Fermi-surface. This is not important under ordinary conditions because the two kinds of Majoranas are at the same energy and at same location. But if there are local excitations which hybridize with one of such pairs and move them to a finite energy from the chemical potential, and not to the other, free Majorana at the Fermi-surface and fermions with zero charge for small finite ${\bf q}$ about the chemical potential result.  The Fermi-surface for a new variety  of fermions  is then protected. The second physical source of the results is that for the mixed-valence problem, the local excitations may be formed from the linear combination of their hybridization and the pairing of the  $f$ and the $d$-fermions with the same spin. At the mixed-valence condition for a single mixed-valence impurity, precisely such a combination as well as hybridization to only one class of one dimensional fermions leads to an analytic solution for a quantum critical point whose results agree with the solution using Wilson renormalization group. 

%To be able to use a pair of diagonal quadratic expressions for fermions that are linear combinations of particles and holes for the kinetic energy of bare fermions near the Fermi-surface, it has been necessary to use  pairing with centers of mass momentum $2{\bf k}_F$. Usually such a pairing is disfavored because of the kinetic energy at such finite momentums. This is not an issue here because one of the partners in pairing are correlated $f$-electrons with effectively infinite mass. It is also worth reiterating that the superfluid density in the problem as solved above is zero.

Using these two general features, together with satisfying the mixed-valence condition for the lattice  leads to  the physical features  giving  single-particle excitations which are fractionalized to  a pair of gapless and spin-less fermion excitations for ${\bf q}$ near the Fermi-wave-vectors and a pair of states similar but not exactly like the hole and particle states in an insulator with a gap. 

The gapless propagating  fermions give specific heat and magneto-oscillations just as ordinary fermions but with 1/2 the amplitude but no Zeeman response. They are dark to any electromagnetic perturbations at long wave-lengths. The gapped excitations are renormalized excitations similar to that of an insulator. The superfluid response is zero. The theory predicts that the magneto-oscillations do not show Zeeman splitting and the magnetization transition to the metallic state at high fields which have been observed.
There are also predictions on the entropy and the single-particle excitation spectra which await experimental results. 

An important unaddressed issue is the effect of impurities which may be significant since the theory relies on
a well-defined Fermi-surface in momentum.

Several other systems feature highly correlated nearly localized  bands near the chemical potential interacting with dispersive bands. They are found in varieties of layered transition metal chalcogenides which have recently been tuned to show heavy-fermi-liquid characteristic \cite{Mak_WSE2_2022} and which may also be tuned to stable mixed-valence conditions. 
%Kagome compounds \cite{Checkelsky_2020}, \cite{Aeppli_2024} with nearly dispersion-free bands close to the chemical potential due to frustration are known with some extraordinary properties. 
 Bands in twisted bi-layer \cite{EAndreiREV2021} and other multi-layer graphene have also been recently mapped \cite{Hu_2023} to dispersion-less bands near wider dispersive bands. In these situations, the new solution for such problems may also be realized.

In the context of such other problems with flat bands near the chemical potential, a general point may be noted. Both the particle-hole and the particle-particle interactions of  states in the flat band with the dispersive band are equally important, just as they are in problems such as the Kondo or mixed-valence impurity problems. 

Intriguing results are obtained in the investigation of a variety of magnetic insulators (see for example the review \cite{Savary16}) which show appear to have shown low energy fermionic thermodynamics. The mixed-valence compounds specifically investigated here present the first unambiguous evidence of such states un-complicated by impurities.
A theory for them may point towards the nature of investigations on the other problems as well. 

Alternative theoretical efforts on this problem may be found in \cite{Baskaran_Majorana, Knolle_Cooper, Erten_Coleman, Chowdhury_Senthil, Fu2018, Fabrizio2023}. A review of the methods and their success in explaining the experiments is not given here. 
  
{\it Acknowledgements}:   Thanks are due to Lu Li, John Singleton  and Suchitra Sebastian for discussion of their experimental data, and for suggestions on the manuscript. Discussions with Thierry Giamarchi and Srinivas Raghu are gratefully acknowledged. l also thank James Analytis, Robert Birgeneau and Joel Moore for arranging my stay at Berkeley, where this work was done.

%\bibliography{REF_APR2023.bib}
%\end{document}
 \bibliography{DarkFermions3.bbl}
 \end{document}